\documentstyle[floats,aps]{revtex}

\begin{document}
\draft

\catcode`\@=11 \catcode`\@=12 
\twocolumn[\hsize\textwidth\columnwidth\hsize\csname@twocolumnfalse\endcsname
\title{A Microscopic Model of Edge States of Fractional 
Quantum Hall Liquid: From Composite Fermions to Calogero-Sutherland Model }

\author{Yue Yu }
\address{Institute of Theoretical Physics, Chinese Academy of  Sciences, 
P. O. Box 2735,
Beijing 100080, China}

\maketitle
\begin{abstract}
 
Based on the composite fermion approach, we derive a microscopic theory 
describing the low-lying edge excitations in the fractional quantum 
Hall liquid with $\nu=\frac{\nu^*}{\tilde\phi\nu^*+1}$. For $\nu^*>0$, it is found that the 
composite fermion model reduces to an SU$(\nu^*)$ Calogero-Sutherland model
in the one-dimensional limit, whereas it is not exact soluble for $\nu^*<0$.
However, the ground states in both cases can be found and the low-lying excitations 
can be shown the chiral Luttinger liquid behaviors since a gap exists between the 
right- and left-moving sectors in each branch of the azimuthal excitations.    

\end{abstract}

]

\section{introduction}

The basic characteristic of the quantum Hall states is the incompressibility
of the two-dimensional (2-d) electron system in a strong perpendicular
magnetic field \cite{Laugh}. While there is a finite energy gap for
particle-hole excitations in the bulk, the low-lying gapless excitations are
located at the edge of the quantum Hall liquid \cite{Halp}. The theoretical
picture of the edge states of fractional quantum Hall (FQH) effect is beyond
the Fermi liquid framework and known as the chiral Luttinger liquid (CLL) 
\cite{Wen,Kane}. Recently, the edge excitations of FQH effect (FQHE) were
studied by several groups numerically \cite{Rez} or in accordance with the
Calogero-Sutherland model (CSM) \cite{ECS} as well as the composite fermion
(CF) picture in the Hartree approximation \cite{CFP}.

In previous works \cite{Yu1}, we have given a microscopic model of the CF
for the edge excitations at the filling factor $\nu=1/m ,~m={\rm odd~integer}
$. It was seen that at $\nu=1/m$ the CF system \cite{Jain} reduces to the
original CSM \cite{CS}. And the low-lying excitations of the edge, then, are
governed by a CLL. We further applied this microscopic theory to analyze the
tunneling experiment \cite{Chang} and got a better fit to the features of
the measured current-temperature curve \cite{Zheng}. However, the CLL is
applied not only to the one channel edge excitation, i.e. , at $\nu=1/m$ but
also to the multi-channel case, for example, at $\nu=\frac{\nu^*}{\tilde\phi%
\nu^*+1}$ for integer $\nu^*$ and even number $\tilde\phi>0$ \cite{Wen}.
Furthermore, a systematic experimental study of the current-voltage (I-V)
characteristic for the electron tunneling between a metal and the edge of a
2-d electron gas at a fractional filling factor $\nu$ shows a continuous
non-Ohmic exponents $\alpha=1/\nu$, i.e., $I\sim V^\alpha$ \cite{Gray}. This
is contradict with the prediction of the CLL theory in which, say, $I\sim
V^3 $ for the primary filling factor being 1/3. The experimentalists even
found that such an I-V characteristic is observed at the filling factor $%
\nu=1/2$ for which the bulk states are compressible. Several authors have
made their efforts in explaining these phenomena \cite{expl,Lee}. Lee and
Wen recently proposed a two-boson model for FQHE regime in which the spin
velocity is much slower than the charge's and then the long time behavior
shows the exponent $\alpha=1/\nu$ while the short time behavior complies
with the Fermi statistics of the electrons \cite{Lee}. They also use the
theory of the half-filling Landau level proposed by Lee \cite{Lee1} to
explain the I-V curve observed in the experiment.

In this work, we would like to generalize our microscopic derivation to the
CLL at $\nu=1/m$ to that at $\nu=\frac{|\nu^*|}{\tilde\phi|\nu^*| \pm 1}$.
(For the bulk state, Cappelli et al have discussed this stable hierachical
quantum Hall liquids from $W(1+\infty)$ minimal model\cite{capp}). 
Tracing the clue of the previous works \cite{Yu1}, it is seen that the edge
theory at $\nu=\frac{\nu^*}{\tilde\phi\nu^* +1}$ is basically described by
the SU$(\nu^*)$ CSM, i.e., the $\nu^*$-branch theory whose low-lying
excitations behaves a charge-`spin separation form where the $\nu^*-1$ spin
branches run in the same direction as the charge branch's. However, it is
not the case for that at $\nu= \frac{|\nu^*|}{\tilde\phi|\nu^*|-1}$ ($%
\nu^*<0 $). The edge theory at these filling factors does not correspond to
an exact soluble model. Fortunately, we could still find a good
approximation ground state wave function and then the $|\nu^*|$ branch
low-lying excitations where the charge branch runs in the opposite direction
to the spin branches. We show that the exclusion statistics between the
branches is described by the $K$ matrices \cite{WenZ}. Thus, using the
bosonization procedure developed in \cite{Wu} the system with the exclusion
statistics matrix $K$ can be taken as the fixed point of the multi-channel
Luttinger liquid. From the radial wave equation of the system, one can show
that the residual magnetic field provides a gap between right- and
left-moving modes in a single branch of these low-lying excitations. This
verifies the CLL at the $T\to 0$ limit. From this microscopic picture, the
spin and charge velocities can be estimated and one finds that $%
v^*_s<<v^*_\rho$. This supports the two-boson theory of Lee and Wen.

This paper is organized as follows: In section II, we present our frame of
work and explain the approximation we used. In section III, we discuss the
solution of the edge Hamiltonian and get the $SU(\nu ^{*})$
Calogero-Sutherland model. In section IV, we discuss the robustness of the
exponents of the CLL. In section V, we prove the chirality of the edge
states. In section VI, we discuss the edge states with filling factors $\nu
=2/3,3/5,...$. The section VII gives our conclusion.

\section{composite fermions at edge}

\subsection{General Formalism}

The two-dimensional interacting electrons which are polarized by a high
magnetic field are governed by the following Hamiltonian 
\begin{eqnarray}
H_{{\rm el}}&=&\sum_{\alpha=1}^N\frac{1}{2m_b}[\vec{p}_\alpha -\frac{e}{c}%
\vec{A}(\vec{r} _\alpha)]^2+\sum_{\alpha<\beta}V(\vec{r}_\alpha-\vec{r}%
_\beta) \\
&+&\sum_\alpha U(\vec{r}_\alpha),  \nonumber  \label{HO}
\end{eqnarray}
where $V(\vec{r})$ is the interaction between electrons. $m_b$ is the band
mass of the electron and $U(\vec{r})$ is the external potential. The
composite particle transformation will bring us to a good starting point to
involve in the FQHE physics as many successful investigations told us \cite
{HLR}. We begin with the CF transformation which reads 
\begin{equation}
\Phi(z_1,...,z_N) = \prod_{\alpha<\beta}\biggl[\frac{z_\alpha-z_\beta} {%
|z_\alpha-z_\beta|}\biggr]^{\tilde\phi} \Psi(z_1,...,z_N),  \label{CBT}
\end{equation}
where $\Phi$ is the electron wave function. The CF consists of an electron
attached by $\tilde\phi$ flux quanta. By using the CF theory, the bulk
behavior of the FQHE has been well-understood \cite{Jain,HLR}. We, now,
would like to study the microscopic theory of the CF edge excitations. The
partition function of the system is given by 
\begin{eqnarray}
Z&=&\sum_{N^e} C^{N^e}_{N}\int_{\partial} d^2z_1....d^2z_{N^e}
\int_{B}d^2z_{N^e+1}...d^2z_N \\
&\times&\biggl(\sum_\delta |\Psi_\delta|^2 e^{-\beta
(E_\delta+E_g)}+\sum_\gamma|\Psi_\gamma |^2e^{-\beta (E_\gamma+E_g)}\biggr),
\nonumber  \label{pf}
\end{eqnarray}
where we have divided the sample into the edge $\partial$ and the bulk $B$. $%
E_g$ is the ground state energy and $E_\delta$ are the low-lying gapless
excitation energies with $\delta$ being the excitation branch index. $%
E_\gamma$ are the gapful excitation energies. At $\nu=1/\tilde\phi$, the
low-lying excitations are everywhere in the sample and we do not consider
this case here. We are interested in the case $\nu=\frac{\nu^*}{\tilde\phi
\nu^*\pm 1}$, where the bulk states are gapful. The low-lying excitations
are confined in the edge of the sample. For convenience, we consider a disc
geometry sample here. The advantage of the CF picture is we have a
manifestation that the FQHE of the electrons in the external field $B$ could
be understood as the IQHE of the CFs in the effective field $B^*$ defined by 
$B^*\nu^*=B\nu$. The energy gap in the bulk is of the order $\hbar \omega_c^*
$ with the effective cyclotron frequency $\omega_c^*=\frac{eB^*}{m^* c}$ ($%
m^*$ is the effective mass of the CF). Hereafter, we use the unit $%
\hbar=e/c=2m^*=1$ except the explicit expressions. By the construction of
the CF, the FQHE of the electrons can be described by the IQHE of the CFs 
\cite{Jain} while the electrons in the field with the filling factor $\nu= 1/%
\tilde\phi$ could be thought as the CFs in a zero effective field. Thus, a
Fermi-liquid like theory could be used \cite{HLR} and we have a set of
CF-type quasiparticles. Applying the single particle picture, which Halperin
used to analyze the edge excitations of the IQHE of the electrons, to the
edge excitations of the CFs, one could have a microscopic theory of the
quasiparticles at the edge. In the low-temperature limit, the domination
states contributing to the partition function are those states that the
lowest Landau level of the CF-type excitations is fully filled in the bulk
but only allow the edge CF-type excitations to be gapless because the gap is
shrinked in the edge due to the sharp edge potential. The other states with
their energy $E_\gamma+E_g$ open a gap at least in the order of $\hbar
\omega^*_c$ to the ground state. In the low-temperature limit, $k_BT\ll
\hbar \omega^*_c$, the effective partition function is 
\begin{eqnarray}
Z&\simeq&\sum_{\delta, N^e}C_N^{N^e} \int_{\partial} d^2z_1...d^2z_{N^e}
|\Psi_{e,\delta}|^2 e^{-\beta (E_\delta(N^e)+E_{g,b})} \\
&=&\sum_{N^e}C_N^{N^e}{\rm Tr_{(edge)}}e^{-\beta (H_e+E_{g,b})},  \nonumber
\label{apf}
\end{eqnarray}
where the trace runs over the low-lying set of the quantum state space for a
fixed $N_e$ and, according to the single particle picture, $\Psi_{e,\delta}$
are the edge many-quasiparticle wave functions . $E_\delta(N^e)$ is the
eigen energy of the edge quasiparticle excitations and $E_{g,b}$ is the bulk
state contribution to the ground state energy. For the disc sample, the edge
quasiparticles are restricted in a circular strip near the boundary with its
width $\delta R(\vec r)\ll R$ while the radius of the disc is $R$. The edge
Hamiltonian of CFs reads 
\begin{eqnarray}
H_e&=&\sum_{i=1}^{N^e}[\vec{p}_i-\vec{A}(\vec{r}_i) +\vec{a}_e(\vec{r}_i)+%
\vec{a}_b(\vec{r}_i)]^2 \\
&+&\sum_{i<j}V(\vec{r}_i-\vec{r}_j)+ \sum_i U_{eff}(\vec{r}_i),  \nonumber
\end{eqnarray}
where the external potential $U_{eff}$ is the effective potential including
the interaction between the edge and bulk particles. The band mass $m_b$ has
been phenomenologically replaced by the CF effective mass. We suppose the
potential is an infinity wall for $r\geq R$. The statistics gauge field $%
\vec{a}$ is given by 
\begin{eqnarray}
\vec{a}_e(\vec{r}_i)&=&\frac{\tilde\phi}{2\pi}\sum_{j\not{=}i} \frac{\hat{z}%
\times (\vec{r}_{i}-\vec{r}_{j})} {|\vec{r}_{i}-\vec{r}_{j}|^2}, \\
\vec{a}_b(\vec{r}_i)&=&\frac{\tilde\phi}{2\pi}\sum_{a} \frac{\hat{z}\times (%
\vec{r}_{i}-\vec{r}_{a})} {|\vec{r}_{i}-\vec{r}_{a}|^2},  \nonumber
\end{eqnarray}
where $a$ is the index of the bulk electrons. Taking the polar coordinate $%
x_i=r_i\cos\varphi_i,~ y_i=r_i\sin\varphi_i$, the vector potential $%
A_\varphi(\vec{r}_i)=\frac{B}{2}r_i$ and $A_r(\vec{ r_i})=0$. In the
mean-field approximation, $a_{r,b}(\vec{r_i})=0$ and $a_ {\varphi,b}(\vec{r}%
_i)=B_{\tilde\phi}r_i/2$. Substituting the polar variations and the vector
potential to $H_e$ while using the mean-field value of $\vec{a}$, one has 
\begin{eqnarray}
H_e&=&\sum_i\biggl[-\frac{\partial^2}{\partial r_i^2}+ (-\frac{i}{r_i}\frac{%
\partial} {\partial \varphi_i}-\frac{B^*}{2}r_i+\frac{m(N_e-1)}{2r_i})^2 \\
&+&\frac{\tilde\phi^2}{4R^2}\sum_i(\sum_{j\not{=}i} \cot\frac{\varphi_{ij}}{2%
})^2  \nonumber \\
&-& \frac{\tilde\phi}{R}\sum_{i<j}\cot\frac{\varphi_{ij}}{2} \cdot i(\frac{%
\partial}{ \partial r_i}- \frac{\partial}{\partial r_j})-\frac{1}{R} \frac{%
\partial}{\partial r_i}\biggr]  \nonumber \\
&+& V+U+O(\delta R/R),  \label{HE}
\end{eqnarray}
where the residual magnetic field $B^{*}=\frac \nu {\nu ^{*}}B$. The
manifestation of CF picture is that the $\nu ^{*}$ denotes the highest
Landau level index of CF in the residual magnetic field. Our central focus
is to solve the many-body problem $H_e\Psi _e(z_1\sigma _1,...,z_{N_e}\sigma
_{N_e})=E\Psi _e(z_1\sigma _1,...,z_{N_e}\sigma _{N_e})$ where $\sigma
_i=1,...,\nu ^{*}$ is the Landau level index which we call spin hereafter.
And the many-body wave function $\Psi _e$ has to be consistent with the bulk
state.

\subsection{Reduce to Calogero-Sutherland Model}

In the previous works \cite{Yu1}, we have presented an example to solve the
problem at $\nu =1/m$, i.e., $\nu ^{*}=1$, and see that the edge ground
states can be directly related to Laughlins wave function in the bulk.
However, it is not applied to a general FQH state with $\nu =\frac{\nu ^{*}}{%
\nu ^{*}\tilde{\phi}+1}$.

It seems that the FQH states with $\nu ^{*}>0$ can be more easily handled
than the states with $\nu ^{*}<0$ as seen below. We start from the easier
one. To the zero order of $V$, we first switch off this interaction. Without
loss of the generality, one takes the trial wave function is of the form 
\begin{eqnarray}
&&\Psi _e(z_1\sigma _1,...,z_{N_e}\sigma _{N_e})  \nonumber \\
&=&\exp \biggl[\frac i2\sum_{i<j}t_{\sigma _i\sigma _j}\frac{r_i-r_j}R\cot 
\frac{\varphi _{ij}}2-\frac A2\frac{r_i-r_j}R(J_{\sigma _i}-J_{\sigma _j})%
\biggr]  \nonumber \\
&&\times f(r_1,...,r_{N_e})\Psi _s(\varphi _1\sigma _1,...,\varphi
_{N_e}\sigma _{N_e}),  \label{TRI}
\end{eqnarray}
where $t_{\sigma _i\sigma _j}$ is a parameter matrix to be determined and so
is $A$. $J_\sigma $ is the spin quantum number. The radial wave function $f$
is symmetric and the azimuthal wave function $\Psi _s$ is anti-symmetric in
the particle exchange. To be consistent with the bulk wave function, the
azimuthal wave function takes its form, 
\begin{eqnarray}
&&\Psi _s(\varphi _1\sigma _1,...,\varphi _{N_e}\sigma
_{N_e})=\prod_{i>j}\phi _{ij}\cdot \prod_k\xi _k^{J_{k_\sigma }},  \nonumber
\\
&&\phi _{ij}=|\xi _i-\xi _j|^{\tilde{\phi}}(\xi _i-\xi _j)^{\delta _{{\sigma
_i}{\sigma _j}}}\exp \{i\frac \pi 2{\rm sgn}(\sigma _i-\sigma _j)\},
\label{AWF}
\end{eqnarray}
where $\xi _i=e^{i\varphi _i}$. In the one-dimensional (1-d) limit, taking $%
\delta R/R\to 0$ in (\ref{HE}) after acting on $\Psi _e$, the Hamiltonian on 
$\Psi _e$ yields $H_{cs}$ on $\Psi _s$ with 
\begin{equation}
H_{cs}=\sum_i(i\frac \partial {\partial x_i}+\frac{B^{*}}2R)^2+\frac{\pi ^2}{%
L^2}\sum_{i<j}\frac{\tilde{\phi}(\tilde{\phi}+P_{\sigma _i\sigma _j})}{[\sin
(\frac{\pi x_{ij}}L)]^2},  \label{HCS}
\end{equation}
where $x_{ij}=x_i-x_j$, $\varphi _i=\frac{2\pi x_i}L$ and $L=2\pi R$ is the
size of the boundary. $P_{\sigma _i\sigma _j}$ is the spin exchange operator 
\cite{HH}. To arrive at (\ref{HCS}), the matrix $t_{\sigma _i\sigma _j}$ is
taken as $\frac 12\delta _{\sigma _i\sigma _j}-\frac 14$ and $A\equiv 1/N_e$%
. The Hamiltonian (\ref{HCS}) is just the SU$(\nu ^{*})$ CSM Hamiltonian
with a constant shift to the momentum operator \cite{SUCS,HH} and the ground
state wave function is given by taking CFs in each branch have the same
number $M$ and $J_1=...=J_{\nu ^{*}}=(M-1)/2$ in $\Psi _s$. In principle, $%
H_e$'s Hilbert space in the 1-d limit can be larger than $H_{cs}$'s.
However, we believe that all interesting azimuthal physics have been
included in $H_{cs}$. Moreover, from the zeros of the ground state wave
function, one can read out the exclusion statistics matrix $K$ \cite{HaldWu} 
\begin{equation}
K_{\sigma \rho }=\tilde{\phi}+\delta _{\sigma \rho }.  \label{Kmatrix}
\end{equation}
Here we see that the mutual exclusion statistics can be different from the
mutual exchange statistics by a Klein factor. It is easy to see that $\nu =%
\frac{\nu ^{*}}{\nu ^{*}\tilde{\phi}+1}=\sum_{\sigma \sigma ^{\prime
}}(K^{-1})_{\sigma \sigma ^{\prime }}$, which shows the consistence between
the edge and the bulk states. Furthermore, the asymptotic Bethe ansatz (ABA)
equation which determines the psudomomentum $n_{i\sigma _i}$ according to
the $K$-matrix (\ref{Kmatrix}). The psudomomentum $n_{i\sigma _i}$ relates
to $J_{\sigma _i}$ in a complicated way and we do not show it explicitly.
Because we have used an SU$(\nu ^{*})$ symmetric form to construct the
azimuthal wave function, the ABA equations are symmetric for the spin
indices, which coincides with the symmetric $K$ matrix.

\section{Non-renormalization of the exponents}

In a work characterizing the Luttinger liquid in terms of the ideal excluson
gas \cite{Wu}, we have bosonized the single component CSM and arrived at the
single branch Luttinger liquid. This procedure can be generalized to
bosonize the SU$(\nu ^{*})$ CSM. The generalization is somewhat trivial but
tedious. We do not present the details for the many branch model here
because the result is just as expected--a $\nu ^{*}$ branch Luttinger liquid 
\cite{HH} with the commutation relations between the neutral edge
excitations $\rho _{\sigma n}$ \cite{Wen} 
\begin{equation}
\lbrack \rho _{\sigma n}^a,\rho _{\sigma ^{\prime }n^{\prime }}^b]=\delta
_{ab}(K^{-1})_{\sigma \sigma ^{\prime }}\frac n{2\pi R}\delta _{n+n^{\prime
}},  \label{Comm}
\end{equation}
where $a,b=L,R$ are the indices of the left- and right-movings. The SU$(\nu
^{*})$ symmetry leads to all excitations have the same velocity. However, we
have to face two problems: i) The edge excitations are chiral which has not
shown in the previous discussion. ii) The robustness of the CLL exponents to
the perturbations. In this section, we focus on the latter by taking the
one-component CLL as an example.

It is well-known that the CSM is an example of one dimensional ideal
excluson gas(IEG) \cite{Wu} with the statistical parameter $m$. And the IEG
is proved to describe the fixed point of the Luttinger liquid. The
bosonization of CSM shows that the low-lying excitations are governed by a $%
c=1$ CFT with the compactified radius $1/\sqrt{m}$ \cite{KY,Wu}.
In this
section, we would like to show that at least some kinds of short-range
interactions between the CFs do not renormalize the topological exponent $g=m
$ under the condition that the scatterings with large momentum transfer
(including  backward scattering and umklapp scattering) are absent because
of the chirality.

To deal with the CSM with interactions, we begin with the asymptotic Bethe
ansatz (ABA) equation \cite{CS}, 
\begin{equation}
k_iL=2\pi I_i+\sum_{j}{^{\prime}}\theta(k_i-k_j),  \label{ABA}
\end{equation}
where $k_i$ is the pseudomomentum of particle $i$, $L$ is the size of the
one dimensional system concerned, and $I_i$ gives the corresponding quantum
number, which is an integer or half-odd. $\theta(k)$ represents the phase
shift of a particle after a single collision with a pseudomomentum transfer
of $k$. It has been proved that ABA equations give exact solutions to the
energy spectrum of the CSM. We assume this approach could be generalized to
the situations of CSM plus some other kind of interaction with force range
shorter than $\frac{1}{r^2}$ potential in the sense of perturbation. This
assumption is justified for the following reasons: First, at the edge of
fractional quantum Hall liquid with $\nu=1/m$, the linear density of the
edge particles can be estimated as 
\begin{equation}
\rho\propto n\times l_B\propto B^{-1/2},
\end{equation}
where $n$ is the average bulk density of the FQH liquid that is fixed and $%
l_B =eB/m^*c$ is the magnetic length corresponding to the magnetic field $B$%
. Under the condition of strong enough magnetic field, the edge particles
can be regarded as a dilute one dimensional gas, where only two body
collisions are important, and the free length between two collisions is long
enough to allow the phase shift to reach its asymptotic value. Secondly,
what we are concerned with is the property of low energy excitations near
the Fermi surface, not the whole precise energy spectrum which can not be
given by ABA. The low energy excitations involve only scattering processes
with small momentum transfer $\Delta k$ (because of chirality, see Sec.IV),
and are determined by the behavior of $\theta(k) $ around $k=0$, which is
dominated by $\theta_{cs}(k)$ (see below). Under the condition of low energy
limit where we let $\Delta k$ approach zero slowly, $\theta(k)$ will become
asymptotically close to $\theta_{cs}(k)$, as a result we can expect ABA
calculations to give asymptotically correct results. Here we implicitly
assume : the low energy spectrum of the CSM varies continuously with respect
to the addition of small perturbation without undertaking any abrupt changes
like a phase transition. This assumption is reasonable, for the low energy
limit of the CSM is the fixed point of Luttinger liquid which is robust
against perturbations. Therefore, what follows from ABA, as we believe, is
credible.  As a matter of fact, for a large class of short-range
interactions, we can expect the ABA works in describing the low-lying
excitations of the system. Indeed, there are several kinds of short-range
interactions whose low-lying excitations are governed by the ABA. An example
of them is the $\delta^{(l)}$-function interaction with $(l)$ representing
the l-th derivative of the $\delta$-function and $l$ being restricted to $l<m
$. The pseduopotentials used by Haldane \cite{psudo} are other examples
because of the vanishing of the expectation value of the pseudopotentials in
the ground state.

To calculate the phase shift, we note the analog of the Schrodinger equation
of the two-body CSM with an additional short-range interaction in the limit $%
L\to \infty$ to the radial equation of a three-dimensional scattering
problem of a centrally symmetric potential. The topological exponent $m$
corresponds to the total angular momentum $l$, i. e. $m=l+1$. The
Schrodinger equation reads 
\begin{equation}
\frac{d^2\psi(x)}{dx^2}+\biggl[(E-V)-\frac{l(l+1)}{x^2}\biggr]\psi(x)=0.
\label{SEC}
\end{equation}
The asymptotic solution of (\ref{SEC}) for $x\gg 0$ is given by 
\begin{equation}
\psi(x)\approx 2\sin(kx-\frac{1}{2}l\pi+\delta_l),
\end{equation}
where $\delta_l$ is the three-dimensional phase shift corresponding to the
scattering potential $V$. In the sense of 1-d scattering, 
\begin{equation}
\theta(k)=\pi(m-1){\rm sgn}(k)-2\delta_l(k).  \label{ps}
\end{equation}
We see that the contribution of $V$ to the phase shift is 
\begin{equation}
\theta_{{\rm reg}}(k)=-2\delta_l(k),
\end{equation}
which is continuous and vanishing at $k=0$ if $V$ is short-ranged (shorter 
than $1/r^2$).

Now, let's make the relation to the macroscopic theory. In terms of the
partition function (\ref{apf}), there is a most probable edge CF number $%
\bar{N}^e$ which is given by $\delta Z/\delta N^e=0$. $\bar{N}^e=\int dx
\rho(x)$ with the edge density $\rho(x)=h(x)\rho_e$ \cite{Wen}. Here $h(x)$
is the edge deformation and $\rho_e$ is the average density of the bulk
electrons. We do not distinguish $\bar{N_e}$ and $N_e$ hereforth if there is
no ambiguity. The low energy properties of the CSM can be obtained from the
ABA equations, 
\begin{eqnarray}
\rho(k)&=&\rho_0(k)-\int\limits_{-k_F}^{k_F}g(k-q)\rho(q) dq \\
\epsilon(k)&=&\epsilon_0(k)-\int\limits_{-k_F}^{k_F}g(k-q)\epsilon(q) dq
\label{tba}
\end{eqnarray}
where $k_F=\pi m N^e_0/L$, 
\begin{equation}
g(k)=\frac{1}{2\pi} \frac{d\theta(k)}{dk},
\end{equation}
$\epsilon_0(k)=k^2-k_F^2$ and $\rho_0(k)=\frac{1}{2\pi} $.

If we consider only the CSM without other interactions ,then we have 
\begin{equation}
\theta_{cs}(k)=\pi(m-1){\rm sgn}(k).  \label{theta}
\end{equation}
Substituting (\ref{theta}) into (\ref{tba}) and after the linearization, we
get 
\begin{eqnarray}
\epsilon_{cs\pm}(k)=\biggl\{{
\begin{array}{ll}
\pm v_+(k\mp k_F), & {\rm if}~ |k|>k_F \\ 
\pm v_-(k\mp k_F), & {\rm if}~ |k|<k_F,
\end{array}
}  \label{LD}
\end{eqnarray}
where
\begin{eqnarray}
v_+&=&\frac{d\epsilon(k)}{dk}\biggl {|}_{k=k_F+0^+}=v_F \\ \nonumber
v_-&=&\frac{d\epsilon(k)}{dk}\biggl {|}_{k=k_F-0^+}=\frac{v_F}{m}
\end{eqnarray}
with $   v_F=2k_F  $ 
and
\begin{eqnarray}   
      \rho_+&=&\rho(k_F+0^+)=\frac{L}{2\pi}  \\ \nonumber
      \rho_-&=&\rho(k_F-0^+)=\frac{L}{2\pi m} 
\end{eqnarray}      
We rewrite the important equations essential to the bosonization for CSM as 
follows
\begin{eqnarray}
       v_+&=&mv_-  \\
       \rho_+&=&m \rho_-.
\label{vrho}
\end{eqnarray}

We rewrite the important equations essential to the bosonization for CSM
as follows 
\begin{eqnarray}
v_+&=&mv_- \\
\rho_+&=&m \rho_-.  \label{vrho}
\end{eqnarray}

A successful bosonization of the theory with the refraction dispersion (\ref
{LD}) has been done by the authors of \cite{Wu} and one shows that the
low-lying excitations of the CSM are controlled by the $c=1$ CFT with its
compactified radius ${\cal R}=1/\sqrt{m}$ \cite{KY}. This implies that the
low-lying states of the CSM have the Luttinger liquid behaviors with the
exponent $g=m$. We will be back to this issue later after we supplies the
chiral constraint and then show that the edge states of FQHE have the CLL
behaviors.

Now, let us see the effects of the interactions. Following our discussion
that leads to the $c=1$ CFT with the compactified radius ${\cal R}=1/\sqrt{m}
$, the relations (\ref{vrho}) are essential. We would like to check if they
are renormalized by the interactions between CFs. Here, we limit our
discussion to the case $m\not{=}1$. We assume the ABA works to describe the
low-lying excitations of the system with an additional short range
interaction, which is consistent with the chirality of the edge excitations.
Differentiating the phase shift (\ref{ps}) with respect to $k$, one has 
\begin{equation}
g(k)=(m-1)\delta (k)+g_{{\rm reg}}(k).
\end{equation}

The continuity of $\theta_{{\rm reg}}$ implies that $g_{{\rm reg}}$ is no
more singular than the $\delta$-function at $k\to 0$. Therefore, we can
prove that the relations (\ref{vrho}) still hold even after we have
introduced a short-range interaction. After differentiating the dressed
energy equation (\ref{tba}) that is assumed holding for the short-range
interaction we are using and in the dilute gas approximation, with respect
to k, we obtain 
\begin{eqnarray}
v_\pm&=&v_0+\int\limits_{-k_F}^{k_F}\epsilon(q)\frac{d}{dq}g(k_F\pm 0^+-q) dq
\\
&=&v_0-\int\limits_{-k_F}^{k_F}\frac{d}{dq}\epsilon(q)g(k_F\pm 0^+-q) dq 
\nonumber \\
&+&\epsilon(k_F)g(k_F\pm 0^+-k_F)  \nonumber \\
&-&\epsilon(-k_F)g(k_F\pm 0^++k_F).  \nonumber
\end{eqnarray}
The definition of $k_F$ ,i.e., $\epsilon(\pm k_F)=0$, leads to 
\begin{eqnarray}
v_+-v_- & = & \int\limits_{-k_F}^{k_F}\frac{d}{dq}\epsilon(q)(m-1)\delta(q-k)
\\
& = & (m-1)v_- .  \nonumber
\end{eqnarray}
Hence 
\begin{equation}
v_+=m v_-.
\end{equation}
The value of $v_\pm$ can be modified by the interactions but the above
relation does not change. Note that if $\epsilon(k_F)=0$ and $%
\epsilon(-k_F)g(k_F\pm 0^++k_F)$ is continuous at $k_F$, the above
conclusion still holds, which will be the case in the CLL derivation of Sec
IV. By performing a similar procedure to $\rho(k) $, we can show (\ref{vrho}%
) for $\rho_\pm$ as well. Therefore one can see that the bosonization
process of the CSM is still applicable in the presence of perturbative
interactions, and the topological exponent $g=m$ is not renormalized by the
short-range interaction. As a result, the compactified radius of the $c=1$
CFT which governs the low-lying excitations of the theory does not change.

Let us give more comments on the conclusion drawn above. This result seems
remarkable at the first sight, when compared with the standard Luttinger
liquid theory, in which we will have the characteristic exponent
renormalized once a short-range perturbative interaction is switched on. In
fact, no inconsistences exist here. In the bosonization of the general
Luttinger liquid, only short range interactions are considered, whose
Fourier transformation $V(k)$ at k=0 possesses no singularity. Even if the
divergence of $V(k)$ as $k$ approaches zero does show up, it is suppressed
by introducing something like a short-range cutoff or a long-range cutoff
which makes the problem concerned more subtle. The exponent so obtained may
be cutoff-dependent. So we can not naively apply it here. In contrast to the
standard approach, the bosonization of the CSM is based on the especially
simple form of the phase shift function of the $1/r^2$ interaction that is
essential to the solution of ABA. The singularity here manifests itself as a
step discontinuity which can be handled easily (no cutoff is needed).
Because of the critical property of the $1/r^2$ interaction, no other
interactions with shorter ranges can alter this discontinuity, which
guarantees the robustness of the bosonization process. In short, the
bosonization of CSM is not so general as the standard one, but it surely
makes a step forward in understanding the low energy physics of nontrivial
interactions.

We emphasize once again that both $1/r^2$ interaction and the chirality
contribute to the robustness of $g=1/m$ when $m>1$. In general, the critical
exponent will be changed by the introduction of other short-range
interactions if the chirality is not present and backward scattering is
allowed. In contrast, in case of $m=1$, where we are actually dealing with a
Fermi liquid, the discontinuity of the phase shift $\theta(k)$ is absent. So
the above argument of robustness fails. An simple example is to consider a $%
\delta$-function interaction. For $m>1$, the short-range divergence of the $%
1/r^2$ potential requires that the wave function vanishes when two particles
approach each other. Hence the $\delta$-function contribution to the phase
shift is completely suppressed in case of $m>1$,while it does show up for $%
m=1$ \cite{YY}. On the occasion of $m=1$, however, the chirality alone
serves as the determinant factor to ensure the non-renormalizability of $g=1$%
, by prohibiting the left-right scattering part of perturbative interactions
from modifying $g$ . Therefore, one can see that the different microscopic
mechanisms for $m>1$ and $m=1$ give the same macroscopic result.

From the above arguments, we see that the topological exponent is invariant
to the perturbations introduced by additional interactions between
particles, if their interaction range is shorter than that of $1/r^2$.
However, the long range nature of Coulomb interaction allows it to dominate
the $1/r^2 $ interaction which gives $g=\nu$. Considering its especially
singular behavior at k=0, we believe that the so called topological index
can no longer survive, if an unscreened Coulomb interaction without any
cutoff really exists. Fortunately, we have several possibilities that will
lead to partial screening of the Coulomb interaction. In real experiments,
the edge electrons actually are not isolated to a wire-like structure. There
are bulk electrons adjacent to them, which can provide mirror charges and
reduce the original Coulomb interaction to a shorter range interaction. What
is more, metal electrodes commonly used in experiments to supply a
confinement potential can also serve as a mirror charges provider. So we
only have to concern ourselves with partly screened Coulomb interaction
instead of the bare one. The effect of short-range interactions has been
discussed in this section.

\section{ chiral Luttinger liquid: the microscopic point of view}

\subsection{Microscopic Derivation of CLL from the Radial Equation}

In the previous sections, we freeze the radial degree of freedom of the edge
particles and see that the azimuthal dynamics is described by the CSM.
However, there are two branches of gapless excitations in the CSM and the
chirality of the edge excitations are not shown. To arrive at the conclusion
of chirality, we take the radial degree of freedom into account. Let us
first make some simplifications before going into details. The interactions
between CFs are assumed to be independent of the radial degree of freedom
because of the small width of the edge. Moreover, we can think of the
interaction between the CFs as consisting  of only the $1/x^2$-type as we
have demonstrated that short-range interactions do not renormalize the
topological exponent $g=m$.

In Sec. II, we take the approximation $r_i\simeq R$ and arrive at the CSM.
Restoring the radial variable, one has the radial eigen equation, which
reads 
\begin{eqnarray}
&&\sum_i\biggl[-\frac{\partial ^2}{\partial r_i^2}+(\frac n{r_i}-\frac{%
|B^{*}|}2r_i)^2\biggr]g(r_1,...,r_{N_e}) \\
&&+[U_{eff}+O(\delta r_i/R)]g(r_1,...,r_{N_e})=Eg(r_1,...,r_{N_e}). 
\nonumber
\end{eqnarray}
where the terms $\frac 1R\frac \partial {\partial r_i}$ have been absorbed
into $g$ by a simple transformation like the multiplication of $e^{-\sum
r_i/R}$. One can see that the radial eigenstate equation can be treated in
the single particle picture except that the pseduomomenta $k=nR$ are related
to one another by the ABA equations(\ref{ABA}). It is reasonable to arrive
at such a result because the interactions between CFs are the functions of $%
\vec{r_i}-\vec{r_j}$ and the radius-dependent part of the interactions is of
order $\delta r/R$. Now we employ the harmonic approximation used by
Halperin in the case of IQHE edge states\cite{Halp}. Let us first turn off
the applied electric field. The radial single particle wave equation in the
stripe approximation reads 
\begin{equation}
-\frac{d^2g}{dy^2}+B^{*2}y^2g=\varepsilon _{+}g,  \label{+}
\end{equation}
for $n>0$ and 
\begin{equation}
-\frac{d^2g}{dy^2}+B^{*2}y^2g+|nB^{*}|g=\varepsilon _{-}g,  \label{-}
\end{equation}
for $n<0$. Here $y=r-R_n$ and 
\begin{equation}
R_n=\sqrt{\frac{2|n|}{|B^{*}|}}.
\end{equation}
Comparing (\ref{+}) with (\ref{-}), we see that the magnetic field separates
the $n<0$ sector from the $n>0$ sector by an energy gap $|n|\hbar \omega
_c^{*}$. Therefore only the $n>0$ (or equivalently, $k>-k_F$) sector needs
to be considered for the low-lying excitations. This is the first sign of
chirality. The harmonic equation (\ref{+}) has its eigenstate energy 
\begin{equation}
\varepsilon _{+,\nu ^{*}}=\hbar \omega _c^{*}((\nu ^{*}-1)+\frac 12),
\end{equation}
if the center of the harmonic potential $R_n\ll R$. This is consistent with
the mean-field approximation to the bulk state because $R_n\ll R$ actually
corresponds to the bulk state of the theory if we recognize that the width
of the harmonic oscillator wave function is about several times the
cyclotron motion radius $R_c^{*}$. Since $R_n$ is the function of $n$, $%
p_n=m^{*}\omega ^{*}R_n$ can be regarded as a momentum-like quantity. The
harmonic oscillator energy for $R-R_n\ll R_c^{*}$ implies that there is no
left-side Fermi point. This provides a necessary condition of the chirality.
To justify the CLL, one should show the existence of gapless excitations on
the right side. It is known that the eigenstate energy at $R_n=R$ is raised
to 
\begin{equation}
\varepsilon _{R,\nu ^{*}}=\hbar \omega _c^{*}\biggl(2(\nu ^{*}-1)+\frac 32%
\biggr),
\end{equation}
because of the vanishing of the wave function at $r=R$. One asks that what
happens if $R_n$ is slightly away from $R$. To see it clearly, we rewrite (%
\ref{+}) as 
\begin{eqnarray}
&&-\frac{d^2g}{d\tilde{y}^2}+B^{*2}\tilde{y}^2g+2B^{*2}(R-R_n)\tilde{y}g 
\nonumber \\
&&+B^{*2}(R_n-R)^2g=\varepsilon _{+}g,  \label{++}
\end{eqnarray}
where $\tilde{y}=r-R$. If $R_n$ is very close to $R$, i. e., $|R-R_n|\leq
R_c^{*}$, one can take the third term as perturbation and a first order
perturbative calculation shows 
\begin{equation}
\delta \varepsilon _{0,\nu ^{*}}=\varepsilon _{+,\nu ^{*}}-\varepsilon
_{R,\nu ^{*}}=v_c^{*}(p_n-p_R)+(p_n-p_R)^2,  \label{e0}
\end{equation}
where $v_c^{*}=2\pi ^{-1/2}l_{B^{*}}\omega _c^{*}$ is of the order of the
cyclotron velocity of the CF corresponding to $B^{*}$ and then of the order
of $v_F$. So, if we take $v_c^{*}\approx v_F$ and note that $p_n-p_R\approx
k-K_0$, the dispersion (\ref{e0}) can be simply rewritten as 
\begin{equation}
\delta \varepsilon _0(k)\approx (k-K_0+k_F)^2-k_F^2,  \label{dr}
\end{equation}
where $k$ is given by the ABA equations (\ref{ABA}). Near the Fermi point $%
K_0$, the dispersion can be linearized as 
\begin{equation}
\delta \varepsilon _0(k)=v_F(k-K_0),
\end{equation}
which implies that there is a right-moving sound wave excitation along the
edge with the sound velocity $v_F$. There is another Fermi point $k=K_0-2k_F$%
, which corresponds to $R_n\approx R-2R_c^{*}$ and is outside of the region
we are considering and in fact belongs to the bulk state. To show that the
above chiral theory has a Luttinger liquid behavior, we extend continuously $%
\delta \varepsilon _0(k)$ to all possible pseudomomenta which obey the ABA (%
\ref{ABA}). Then, the equation (\ref{dr}) and (\ref{ABA}) mean that the
system is an IEG \cite{Wu}. The problem can be solved by using a
bosonization procedure developed in ref. \cite{Wu}. The edge excitations can
be obtained by considering only the properties of such an IEG system near $%
k\sim K_0$. Consequently, the low-lying excitations of the theory are
controlled by the $c=1$ CFT with its compactified radius ${\cal R}=1/\sqrt{m}
$ as we point out in the discussion of the CSM in Sec. III. However, the
relevant excitations of the edge states include only the right-moving
branch. In other words, the edge states are chiral and the sound wave
excitations correspond to the non-zero modes of the right-moving sector of
the $c=1$ CFT. There are two other kinds of edge excitations which
correspond to the particle additions to the ground state and the current
excitations along the edge respectively. The velocity relations of these
excitations are given by \cite{Wu} 
\begin{equation}
v_M=mv_F,v_J=v_F/m,v_F=\sqrt{v_Mv_J}.  \label{vr}
\end{equation}
The relations resemble those of Haldane's Luttinger liquid if one identifies 
$m$ with the characteristic parameter $e^{-2\varphi }$ in the Luttinger
liquid theory \cite{Hald}. These observations are crucial to the conclusion
that the edge states are controlled by the $c=1$ CFT with its compactified
radius ${\cal R}=1/\sqrt{m}$.

To arrive at the effective theory of CLL, let us perform the following
bosonization procedure.

According to (\ref{dr}) and (\ref{ABA}), the edge excitations with the
pseudomomentum $k$ have their dressed energy 
\begin{eqnarray}
\varepsilon(k)=\biggl\{{
\begin{array}{ll}
(k^2-k^2_F)/m, & |k|<k_F, \\ 
k^2-k^2_F, & |k|>k_F.
\end{array}
}
\end{eqnarray}
Here we have made a translation $k\to k+K_0-k_F$. The linearization
approximation of the dressed energy near $k\sim\pm k_F$ is given by (\ref{LD}%
). In terms of the linearized dressed energy, we obtain a free fermion-like
representation of the theory and then can easily bosonize it \cite{Wu}. The
Fourier transformation of the right-moving density operator is given by 
\begin{eqnarray}
\rho_q^{(+)}&=&\displaystyle \sum_{k>k_F}:c^\dagger_{k-q}c_k: + %
\displaystyle \sum_{k<k_F-m q} :c^\dagger_{k+mq}c_k: \\
& &+ \displaystyle \sum_{k_F-m q< k < k_F}:c^\dagger_{\frac{k-k_F} {m}%
+k_F+q}c_k:,
\end{eqnarray}
for $q>0$ is the sound wave vector. Here $c_k$ is a fermion annihilation
operator. And a similar $\rho_q^{(-)}$ can be defined near $k\sim -k_F$. The
bosonized Hamiltonian is given by 
\begin{equation}
H_B=v_F\{ \sum_{q>0}q(b_q^\dagger b_q +\tilde{b}_q^\dagger \tilde{b}_q) +%
\frac{1}{2}\frac{\pi}{L}[m M^2 +\frac{1}{m} J^2] \}.  \label{bosonH}
\end{equation}
Thus, we have a current algebra like 
\begin{equation}
[\rho_q^{(\pm)},\rho_q^{(\pm)\dagger}]= \frac{L}{2\pi} q\,
\delta_{q,q^{\prime}},~~ [H_B,\rho_q^{(\pm)}]= \pm v_Fq\rho_q^{(\pm)}.
\label{density}
\end{equation}

In the coordinate-space formulation, the normalized density field $\rho(x)$
is given by $\rho(x)=\rho_R(x)+\rho_L(x)$: 
\begin{equation}
\rho_L(x)=\frac{M}{2L}+ \displaystyle\sum_{q>0}\sqrt{q/2\pi Lm}
(e^{iqx}b_q+e^{-iqx}b_q^\dagger),  \label{rhofield}
\end{equation}
and $\rho_R(x)$ is similarly constructed from $\tilde{b}_q$ and $\tilde{b}%
_q^\dagger$. Here ${b}_q=\sqrt{2\pi/qL}{\rho}_q^{(+)\dagger}$ and so on. The
boson field $\phi(x)$, which is conjugated to $\rho(x)$ and satisfies $%
[\phi(x),\rho(x^{\prime})]=i\delta(x-x^{\prime})$, is $\phi(x)=\phi_R(x)+%
\phi_L(x)$ with 
\[
\phi_L(x)= \frac{\phi_{0,L}}{2}+\frac{\pi Jx}{2L}+i \displaystyle \sum_{q>0}%
\sqrt{\pi m/ 2qL}(e^{iqx}b_q-e^{-iqx}b_q^\dagger), 
\]
and a similar $\phi_{L}(x)$. Here $M$ and $J$ are operators with integer
eigenvalues, and $\phi_0=\phi_{l0}+\phi_{r0}$ is an angular variable
conjugated to M: $[\phi_0,M]=i$. The Hamiltonian (\ref{bosonH}) becomes 
\begin{equation}
H_B =\frac{v_F}{2\pi} \int_0^Ldx\; [\Pi(x)^2+(\partial_xX(x))^2],
\label{fieldH}
\end{equation}
where $\Pi(x)=\pi m^{1/2}\rho(x)$ and $X(x)=m^{-1/2}\phi(x)$. With $%
X(x,t)=e^{iHt}X(x)e^{-iHt}$, the Lagrangian density reads 
\begin{equation}
{\cal L}=\frac{v_F}{2\pi}\,\partial_\alpha X(x,t)\,\partial^\alpha X(x,t).
\end{equation}
We recognize that ${\cal L}$ is the Lagrangian of a $c=1$ CFT. Since $%
\phi_{0}$ is an angular variable, there is a hidden invariance in the theory
under $\phi\to\phi+2\pi$. The field $X$ is thus said to be ``compactified''
on a circle, with a radius that is determined by the exclusion statistics: 
\begin{equation}
X\sim X+2\pi {\cal R},~~~ {\cal R}^2=1/m.
\end{equation}
States $V[X]|0\rangle$ or operators $V[X]$ are allowed only if they respect
this invariance, so quantum numbers of quasiparticles are strongly
constrained.

In the present case, only the right-moving sector is relevant. So, we have
an `almost' chiral edge state theory whose sound wave excitation is chiral
but there are charge leakages between the bulk and the edge. The leakages
are reflected in the zero-mode particle number and current excitations \cite
{foot}. In this almost chiral theory, the charge-one fermion operator is
defined by 
\begin{equation}
\Psi_R^\dagger(x)=\sum_{l=-\infty}^\infty\exp(i2(l+\frac{1}{2}m)\theta_R(x))
\exp(i\phi_R(x)),
\end{equation}
where 
\begin{equation}
\theta_R(x)=\pi\int_{-\infty}^x \rho_R(x^{\prime})dx^{\prime}.
\end{equation}
The correlation function, then, can be calculated \cite{EX} 
\begin{eqnarray}
<\Psi_R^\dagger(x,t)\Psi_R(0,0)>&=&\sum_{l=-\infty}^{\infty} C_l\biggl(\frac{%
1}{x-v_Ft})^{(l+m)^2/m}  \nonumber \\
&&\exp(i(2\pi(l+\frac{1}{2})x/L)).
\end{eqnarray}
The $l=0$ sector recovers Wens result \cite{Wen}. In other words, the
present theory justifies microscopically Wens suggestion of CLL of the FQHE
edge states .

\section{many branch theory}

To see the chirality in many branch theory, we can apply the discussion in
the previous section to the $\nu^*$-branch. The eigenenergy of the single
wave function $g_+(y)=g(r_n-R)$ with $n>0$ is $|n|\hbar\omega^*_c$ lower
than that of $g_-(y)=g(r_n-R)$ with $n<0$\cite{Yu1}.Therefore, the states
with the negative $n$ are ranged out of the low-lying state sector. This
implies that only the left-moving mode in the azimuthal dynamics belongs to
the low-lying excitation sector. The width of the wave function $g_+$ is
several times of $R_c^*$, the cyclotron radius of the CF in the effective
field. For the highest spin $\nu^*$, one can see that the eigenenergy of $%
g_+(0)$ ($r_{n_{\nu^*}} =R$) is $\varepsilon_{R\nu^*}=(2(\nu^*-1)+\frac{3}{2}%
) \hbar\omega^*_c$ since the wave function vanishes at $r=R$ while $%
\varepsilon_{n_{\nu^*}}=((\nu^*-1)+ \frac{1}{2})\hbar\omega^*_c$ for $%
R-r_n\gg R_c^*$, a harmonic oscillator energy and coinciding with the mean
field theory applied to the bulk states. The gapless excitations appear when 
$|R-r_n|\sim R_c^*$ . Using a perturbative calculation, one has the bare
excitation energy is $\varepsilon_0=\varepsilon_{n_{\nu^*}}-\varepsilon_{R%
\nu^*} \sim v_F\frac{eB^*}{c}(r_n-R)$ with the Fermi velocity $v_F=\frac{%
\pi\hbar\rho_0}{m^*}(\tilde\phi+\nu^*)$ \cite{Frad}. So, we see that the
Fermi velocity then the CF effective mass $m^*$ are determined by the slope
of the edge spectrum.

The SU$(\nu ^{*})$ symmetry forces all other branches of the edge
excitations have the same velocity both in their magnitue and direction as
the $\nu ^{*}$-th branch. This is contradict to the recent edge tunneling
experiment \cite{Gray}. Lee and Wen \cite{Lee} argued that this
inconsistency could be dispelled if the spin mode velocities are much
smaller than the charge's. ( Charge mode is given by $\rho _c\propto \sum
\rho _\sigma $ and so on.) In our model, there are two factors to change the
sound velocity $v_c$ which equals to $v_F$ before considering those factors.
First, we have taken the effective potential as an infinite wall to simplify
our model. In real samples, this potential is also smooth. This implies that
the real edge spectrum is much flatter than that in the infinite wall
potential case and then the real sound velocity $\tilde{v}_c$ is much
smaller than $v_F$. Equivalently, the CF effective mass gains due to the
smoothness of the edge potential. The other factor to affect the sound
velocity is the interaction $V$ which violates the SU$(\nu ^{*})$ symmetry.
The interaction is of the form $\rho _\sigma V_{\sigma \sigma ^{\prime
}}\rho _{\sigma ^{\prime }}$. In the bosonic form, it can be rewritten as $%
\rho _cV_\rho \rho _c+\rho _sV_{sc}\rho _c+\rho _sV_{ss^{\prime }}\rho
_{s^{\prime }}$ where $\rho _c$ is the charge density wave and $\rho _s$ are
the $\nu ^{*}-1$ branches of spin density wave. All the interactions stem
from the electron-electron interaction, which yields $V_{ss^{\prime
}},V_{sc}<<V_\rho $. We can assume $V_{ss^{\prime }}=V_{sc}=0$. Therefore,
only the charge density wave velocity is renormalized by the interaction $%
V_\rho $. Finally, we have the renormalized velocities 
\begin{equation}
v_\rho ^{*}=\tilde{v}_c+V_\rho ,~~~~v_s^{*}=\tilde{v}_c.
\end{equation}
We see that the smooth edge potential suppresses the spin wave velocity
whereas the interaction $V_\rho $ is not changed. Here we consider the
interaction to be short range. In reality, it may be an unscreened Coulomb
interaction. It is reasonable to take $V_\rho $ in the order of the average
Coulomb interaction, $\frac{2\pi e^2l_B}\epsilon $, which is almost the
inverse of the bulk CF effective mass. However, $\tilde{v}_c$ is
proportional to the inverse of the edge CF effective mass. As the edge
potential is flatted, the edge CF effective mass increases a lot such that $%
\tilde{v}_c<<V_\rho $. Therefore, $v_\rho ^{*}\approx V_\rho >>v_s^{*}$
which is just what Lee and Wen predicted \cite{Lee}.

\section{$\nu=\frac{|\nu^*|}{|\nu^*|\tilde\phi-1}$ edge state}

Turn to much subtle problem at $\nu^*<0$, i.e., $\nu=\frac{|\nu^*|}{|\nu^*|%
\tilde\phi-1}$ . To solve the problem, we make an anyon transformation for
the edge CF with the statistical parameter $b\delta_{\sigma\sigma^{\prime}}$
where b is a real number which is given by a solution of equation $(\tilde%
\phi-1)b^2+2\tilde\phi+2=0$. Although the 1-d limit model is still not
soluble even with this transformation, we may attract the low-lying
excitation sector by using a trial wave function. A trial wave function
enlightened by the bulk wave function may be taken of the form 
\begin{eqnarray}
&&\Psi_e(z_1\sigma_1,...,z_{N_e}\sigma_{N_e})  \nonumber \\
&&=\exp\biggl[\frac{i}{2} \sum_{i<j}s_{\sigma_i\sigma_j}\frac{r_i-r_j}{R}%
\cot \frac{\varphi _{ij}}{2}-\frac{A}{2}\frac{r_i-r_j}{R} (J_{\sigma_i}-J_{%
\sigma_j})\biggr]  \nonumber \\
&&\times f(r_1,...,r_{N_e})
\Psi_s(\varphi_1\sigma_1,...,\varphi_{N_e}\sigma_{N_e}),  \label{TRI1}
\end{eqnarray}
where $s_{\sigma_i\sigma_j}=-\frac{1}{4}+c\delta_{\sigma_i\sigma_j}$ with $%
c=(\frac{1}{4\tilde\phi}-\frac{1}{2})b-\frac{1}{2}$. The azimuthal wave
function $\Psi_s$ can be set according to the $K$ matrix of the bulk state
and reads 
\begin{eqnarray}
&&\Psi_s(\varphi_1\sigma_1,...,\varphi_{N_e}\sigma_{N_e})
=\prod_{i>j}\phi_{ij}\cdot\prod_k \xi_k^{J_{k_\sigma}}  \nonumber \\
&&\phi_{ij}=|\xi_i-\xi_j|^{\tilde\phi}(\xi_i-\xi_j)^{-\delta_{{\sigma_i} {%
\sigma_j}}}  \nonumber \\
&&\times\exp\{i\frac{\pi}{2}[{\rm sgn}(\sigma_i-\sigma_j) +b{\rm sgn}%
(i-j)\delta_{\sigma_i\sigma_j}]\}.  \label{AWF1}
\end{eqnarray}
This wave function is indeed an eigen wave function of the anyon-transformed
Hamiltonian if $\delta R/R\to 0$. Taking a suitable set of the quantum
numbers as that for $\nu^*>0$, we have the ground state wave function. The
important matter is that the exclusion statistics of the azimuthal wave
function is given by the expected bulk $K$ matrix 
\begin{equation}
K_{\sigma\sigma^{\prime}}=\tilde\phi-\delta_{\sigma\sigma^{\prime}}.
\end{equation}
According to this exclusion statistics matrix and our bosonization approach,
we can finally arrive the CLL theory in which the charge mode travels in the
opposite direction than the $\nu^*-1$ spin-modes, which in the clean edge ,
leads to the absence of the edge equilibration . However, the effective
potential $U_{eff}$ includes all possible external potential. Of them, a
most relevant one is the random impurity potential. Kane et al have shown
that this random potential drives the edge to a stable fixed point and
restores the edge equilibration \cite{Kane}.

\section{conclusion}

In conclusion, we have proposed a microscopic model of edge excitations for
FQHE at $\nu=\frac{|\nu^*|}{|\nu^*|\tilde\phi\pm 1}$. The SU$(\nu^*)$ CSM is
a good candidate for the edge states for $\nu^*$ while there is no exact
soluble counterpart for $\nu^*<0$. The low-lying excitations for both $%
\nu^*>0$ and $\nu^*<0$, however, are proven to be described by the CLL. We
also argued that the two-boson theory of Lee-Wen is valid to explain the
experiments by Grayson at el in FQH regime while we did not discuss the
explanation to the tunneling experimental result for the other filling
factors such as $\nu=1/2$ and in the non-FQH regime.

The author is very grateful to Z. Y. Zhu and W. J. Zheng for collaboration..
He thanks T. Xiang and G. M. Zhang for useful discussions. This work was
supported in part by the NSF of China.


\vspace{-0.1in}

\begin{references}
\vspace{-0.4in}

\bibitem{Laugh}  R. B. Laughlin, Phys. Rev. Lett. {\bf 50}, 1395 (1983).

\bibitem{Halp}  B. I. Halperin, Phys. Rev. B {\bf 25}, 2185 (1982).

\bibitem{Wen}  X. G. Wen, Phys. Rev. Lett. {\bf 64}, 2206 (1990); Phys. Rev.
B {\bf 41}, 12838 (1990).

\bibitem{Kane}  C. L. Kane, M. P. A. Fisher and J. Polchinski, Phys. Rev.
Lett. {\bf 72}, 4129 (1994). C. L. Kane and M. P. A. Fisher, Phys. Rev. B 
{\bf 51}, 13449 (1995).

\bibitem{Rez}  E. H. Rezayi and F. D. M. Haldane, Phys. Rev. B {\bf 50},
6924 (1994).

\bibitem{ECS}  F. D. M. Haldane, Bull. Am. Phys. Soc. {\bf 37}, 164(1992);
Phys. Rev. Lett. {\bf 67}, 937 (1991); S. Mitra and A. H. MacDonald, Phys.
Rev. B {\bf 48}, 2005 (1993); P. J. Forrester and B. Jancovici, J. Phys.
(Pairs) {\bf 45}, L583 (1994); N. Kawakami, Phys. Rev. Lett. {\bf 71},
275(1993); A. D. Veigy and S. Ouvry. Phys. Rev. Lett. {\bf 72}, 121(1994).

\bibitem{CFP}  L. Brey, Phys. Rev. B {\bf 50}, 11861 (1994); D. B.
Chkolvskii, Phys. Rev. B {\bf 51}, 9895 (1995).

\bibitem{Yu1}  Y. Yu and Z. Y. Zhu, Commun. Theor. Phys. {\bf 29}, 351
(1998). Y. Yu, W. J. Zheng and Z. Y. Zhu, Phys. Rev. B {\bf 56}, 13279
(1997).

\bibitem{Jain}  J. K. Jain, Phys. Rev. B {\bf 41}, 7653 (1990) and
references therein.

\bibitem{CS}  F. Calogero, J. Math. Phys. {\bf 10} 2197 (1967). B.
Sutherland, J. Math. Phys. {\bf 12}, 246, 251 (1971).

\bibitem{Chang}  A. M. Chang, . L. N. Pfeiffer and K. W. West, Phys. Rev.
Lett. {\bf 77}, 2538 (1996).

\bibitem{Zheng}  W. J. Zheng and Y. Yu, Phys. Rev. Lett. {\bf 79}, 3242
(1997).

\bibitem{Gray}  M. Grayson, D. C. Tsui, L. N. Pfeiffer and K. W. West and A. M. Chang,
 Phys. Rev. Lett. {\bf 80}, 1062 (1998); 
A.M. Chang, L.N. Pfeiffer, and K.W. West, Phys. Rev. Lett. {\bf 77}, 
2538 (1996); A.M. Chang, L.N. Pfeiffer, and K.W. West, 
Physica {\bf B249-251}, 383 (1998)

\bibitem{expl}  A. V. Shytov, L. S. Levitov and B. I. Halperin, Phys. Rev.
Lett. {\bf 80}, 141 (1998). S. Conti and G. Vignale, cond-mat/9801318. U.
Zulicke and A. H. Macdonald, cond-mat/9802019.

\bibitem{Lee}  D. H. Lee and X. G. Wen, cond-mat/9809160; K. Imura,
cond-mat/9812400.

\bibitem{Lee1}  D. H. Lee, Phys. Rev. Lett. {\bf 89}, 4745 (1998).

\bibitem{capp} Andrea Cappelli, Carlo A. Trugenberger 
, Guillermo R.
Zemba, Nucl.Phys.B {\bf 448}, 470,1995;Andrea Cappelli, 
Guillermo R. Zemba, Nucl.Phys.B {\bf 540},610-638,1999; 
Andrea Cappelli, Carlos Mendez,
Jorge M. Simonin, Guillermo R. Zemba, cond-mat/9806238 
to be published in Phys. Rev. B (1999).

\bibitem{WenZ}  X. G. Wen and A. Zee, Phys. Rev. B {\bf 46 }, 2290 (1992).

\bibitem{Wu}  Y. S. Wu and Y. Yu. Phys. Rev. Lett. {\bf 75}, 890 (1995).

\bibitem{HLR}  B. I. Halperin, P. A. Lee and N. Read, Phys. Rev B {\bf 47},
7312 (1993). V. Kalmeyer and S. C. Zhang, Phys. Rev. B {\bf 46}, 9889 (1992).

\bibitem{RCF}  R. Shankar and G. Murthy, Phys. Rev. Lett.{\bf 79}, 4437
(1997). V. Pasquier and F. D. M. Haldane, Nucl. Phys. B {\bf 516},
719(1998). N. Read, cond-mat/9804294; A. Stern, B. I. Halperin, F. von Oppen
and S. H. Simon, cond-mat/9812135.

\bibitem{HH}  Z. N. C. Ha and F. D. M. Haldane, Phys. Rev. B {\bf 46}, 9359
(1992).

\bibitem{SUCS}  Y. Kuramoto and H. Yokoyama, Phys. Rev. Lett. {\bf 67}, 1338
(1991); N. Kawakami, Phys. Rev. B {\bf 46}, 1005 (1992); K. Hikami and M.
Wadati, J. Phys. Soc. Jpn. {\bf 62}, 469 (1993); J. A. Minahan and A. P.
Polychronakos, Phys. Lett. B {\bf 302} 265 (1993).

\bibitem{HaldWu}  F. D. M. Haldane, Phys. Rev. Lett. {\bf 67}, 937 (1991).
Y. S. Wu, Phys. Rev. Lett. {\bf 73}, 922 (1994).

\bibitem{Frad}  We have taken the point that the boundary of the sample is
the Fermi surface of the system (E. Fardkin, {\it Field Theories of
Condensed Matter Systems}, (Addison-Wesley, Redwood City, 1991).)

\bibitem{KY}  N. Kawakami and S.-K. Yang, Phys. Rev. Lett. 67, 2493 (1991).

\bibitem{psudo}  F.D. M Haldane, Phy. Rev. Lett. {\bf 51},605 (1983).

\bibitem{Hald}  F. D. M. Haldane, J. Phys. {\bf C 14}, 2585 (1981).

\bibitem{foot}  In an isolated chiral theory coupled to the external fields
the charge is not conserved because it is known that the theory is anomalous
and not physical \cite{EX}.

\bibitem{EX}  See , e. g. Z.N.C. Ha, Nucl. Phys. {\bf 435}, 604 (1995).
\end{references}
\end{document}